\documentclass[prl,amsmath,amssymb,twocolumn, showpacs, superscriptaddress,10pt]{revtex4}

\usepackage{amsmath}
\usepackage{hyperref}
\usepackage{graphicx}
\usepackage{amsfonts}
\usepackage{amsthm}
\usepackage{cases}
\usepackage{bm}
\usepackage{ulem}

\usepackage{color}
\definecolor{Blue}{rgb}{0.00, 0.00, 1.00}
\definecolor{Red}{rgb}{1.00, 0.00, 0.00}

\hypersetup{
    colorlinks=true,       
    linkcolor=red,          
    citecolor=blue,        
    filecolor=magenta,      
    urlcolor=cyan           
}

\newcommand{\be}{\begin{equation}}
\newcommand{\ee}{\end{equation}}
\newcommand{\bea}{\begin{eqnarray}}
\newcommand{\eea}{\end{eqnarray}}
\newcommand{\ba}{\begin{align}}
\newcommand{\ea}{\end{align}}

\newcommand\underrel[2]{\mathrel{\mathop{#2}\limits_{#1}}}

\allowdisplaybreaks

\usepackage{graphicx}
\newcommand \du {\dot{u}}

\begin{document}

\title{Universal Scaling of the Velocity Field in Crack Front Propagation}

\author{Cl{\'e}ment Le Priol}
\affiliation{CNRS - Laboratoire de Physique de l'Ecole Normale Sup\'erieure, 24 rue Lhomond, 75231 Paris Cedex, France}
\author{Julien Chopin}
\affiliation{Instituto de F{\'i}sica, Universidade Federal da Bahia, Salvador-BA, 40170-115, Brazil}
\author{Pierre Le Doussal}
\affiliation{CNRS - Laboratoire de Physique de l'Ecole Normale Sup\'erieure, 24 rue Lhomond, 75231 Paris Cedex, France}
\author{Laurent Ponson}
\affiliation{Institut Jean le Rond d'Alembert, Sorbonne Universit\'e, 75252 Paris Cedex 05, France}
\author{Alberto Rosso}
\affiliation{LPTMS, CNRS, Universit\'e Paris-Sud, Universit\'e Paris-Saclay, 91405 Orsay, France}

\date{\today}

\begin{abstract}
The propagation of a crack front in disordered materials is jerky and characterized by bursts of activity, called avalanches. 
These phenomena are the manifestation of an out-of-equilibrium phase transition originated by the disorder. 
As a result avalanches display universal scalings which are, however, difficult to characterize in experiments at finite drive.
Here, we show that the correlation functions of the velocity field along the front allow us to extract the critical exponents of the transition and to identify the universality class of the system.
We employ these correlations to characterize the universal behavior of the transition in simulations and in an experiment of crack propagation.
This analysis is robust, efficient, and can be extended to all systems displaying avalanche dynamics. 
\end{abstract}


\maketitle

The presence of disorder is often at the origin of physical behaviors that are not observed in pure systems. 
In particular, under a slow drive, a disordered system does not respond smoothly but is characterized by quick and large rearrangements called \textit{avalanches} followed by long quiescent periods. 
The earthquakes in tectonic dynamics \citep{fisher1997, fisher1998, jagla2014}, the plastic rearrangements in amorphous materials \citep{lin2014, nicolas2018} or the Barkhausen noise in soft magnets \cite{zapperi1998, laurson2013, durin2016} are examples of such avalanches.
If the drive is very slow, avalanches are triggered one by one : the system is driven to a first instability and then evolves freely until it stops. 
In this \textit{quasi-static limit} one can measure the size and the duration of each avalanche. Their statistics are scale-free on many decades, revealing a critical behavior independent of many microscopic details. 

This behavior is well established for earthquakes and Barkhausen noise where events are well separated in time. 
However in most experimental systems, the driving velocity is finite, so that a subsequent avalanche is often triggered before the previous one stops.
One of the standard propositions to define avalanches is to threshold the {\it global} velocity signal. 
However this kind of analysis raises important issues :
If the threshold is too large, a single avalanche may be interpreted as a series of seemingly distinct events,  
while if it is too small subsequent avalanches can be merged into a single event~\cite{janicevic2016, bares2013, bares2019a, bares2019}.
Hence disentangling avalanches becomes nearly impossible and accurately measuring critical exponents is then particularly challenging. 
An alternative method to define avalanches is to threshold the local velocity signal in order to establish the state (quiescent or active) of each point in the system~\cite{maloy2006}. The issue is then to decide wether two active regions separated in space and/or time do belong to the same avalanche or not. The latter problem is particularly severe when studying the propagation of cracks~\citep{tanguy1998, bonamy2008, bonamy2011, ponson2016} and wetting fronts \cite{roux2003, moulinet2004, ledoussal2009a} in disordered materials. 
In these systems the interactions are proven to be long-ranged \cite{rice1985, joanny1984} and quasi-static avalanches are spatially disconnected objects \cite{laurson2010}. Hence reconstructing avalanches from the resulting map of activity clusters remains very difficult, as for large systems there are active points at any time. 

\begin{figure}[t]
\includegraphics[width=1 \linewidth]{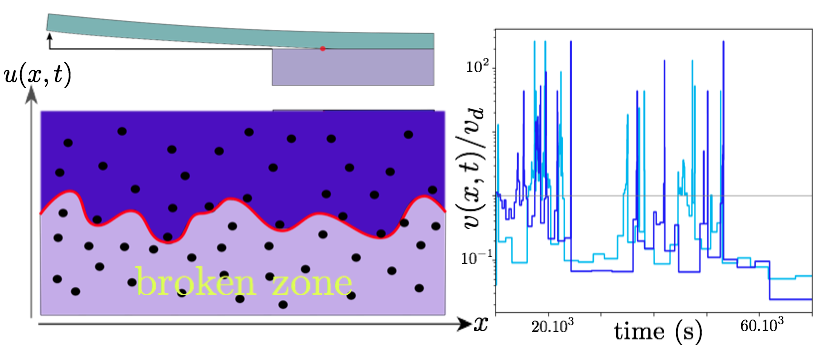}
\caption{\textbf{Left:} Sketch of the experimental setup of planar crack propagation. Profile view (top left): 
a plexiglas plate is detached from a thick silicone substrate at fixed velocity. 
Top view (bottom left) : the crack front (red line) separates the broken region from the unbroken one. Defects are dots of diameter $d_0=100~ \mu\mathrm{m}$.
\textbf{Right:} Local velocity of two points which are $3\, d_0$ apart along the front. The grey line corresponds to the average speed $v_d$. Each signal is intermittent and the two signals display clear correlations. \label{Fig:Intro}}
\end{figure}

In this Letter we develop an alternative strategy. We show that the study of the space and time correlations of the local velocity field, a quantity which is experimentally accessible, allows to capture the universal features of the dynamics without any arbitrariness nor any tunable parameter, even at finite driving speed.
We propose and characterize the scaling forms of these functions and show how they relate with the critical exponents of the avalanche dynamics and with the range of the interactions in the system.
Our predictions are tested on numerical simulations and experimental data of crack propagation,
but are expected to hold for all systems displaying avalanche dynamics but driven at finite velocity.

In Fig.~\ref{Fig:Intro} we show a sketch of the crack front where $u(x,t)$ is the front position at point
$x$ and time $t$.  
Its equation of motion in adimensional units writes (see Eq.(11) of the Supplemental Material \cite{SupMat}
and also \cite{griffith1920, freund1990, demery2014, basu2019}) :
\begin{equation}\label{eq:Movement equation}
\frac{v(x,t)}{\mu} = f + \eta\left(x, u(x,t)\right) + \frac{1}{\pi} \int \frac{u(x',t) - u(x,t)}{|x'-x|^{1+\alpha}} dx' \, 
\end{equation}
with $v(x,t) = \partial_t u(x,t)$.
The mobility $\mu$ has the dimension of a velocity. In an ideal elastic material it coincides with the Rayleigh velocity $c_R$ but it is in general much smaller \cite{SupMat}. 
The first term is the adimensional force $f$ which drives the crack propagation, $\eta(x,u)$ is the normalized toughness fluctuations 
and the last term accounts for the elasticity along the interface. In general this interaction is long-range 
with $0 < \alpha \leq 2$ ($\alpha = 2$ corresponds to short-range elasticity).
In particular for the crack \cite{gao1989} and the wetting fronts \cite{joanny1984} it was shown that $\alpha=1$.

The competition between elasticity and disorder in \eqref{eq:Movement equation} is at the origin of a second order dynamical phase transition called depinning \cite{narayan1993, leschhorn1997}.
The force $f$ is the control parameter and the velocity $v$ is the order parameter which vanishes at a critical force $f_c$. 
In analogy with equilibrium phase transitions, two independent exponents can be defined :
the exponent $\beta$ associated to the order parameter, $v \sim \left(f-f_c\right)^{\beta}$ 
and the roughness exponent $\zeta$ associated to the fluctuations of the front position, 
$ \left\langle \left( u(x,t) - u(0,t) \right)^2 \right\rangle \sim x^{2\zeta} \; $ ; the brackets $\langle ... \rangle$ denote the average over different realizations of the disorder.

Below $f_c$ the velocity is zero, but a local perturbation can induce an extended reorganization of the front, 
\textit{the avalanche}, up to a scale $\xi \sim |f-f_c|^{-\nu}$, the divergent correlation length of the transition. 
Symmetries and dimensional analysis allow to link all the exponents of the avalanche statistics (size, duration, ...) to $\beta$ and $\zeta$.
In particular, the statistical tilt symmetry ensures the scaling relation $\nu = 1/(\alpha - \zeta)$.

In the moving phase it is customary to work with a fixed driving velocity $v_d$ instead of a fixed force $f$. 
In practice, this is achieved by replacing $f$ with a parabolic potential of curvature $m^2$ moving at velocity $v_d$ : 
$f \rightarrow m^2(v_d t - u(x,t))$ \cite{SupMat}.
When $v_d$ is small, the local velocity field $v(x,t)$ along the front displays 
two features which are a clear manifestation of the presence of avalanches~:
(i) it is very intermittent in time, i.e. it is either large of order $v^{\max} \gg v_d$ or almost zero and (ii) it displays strong correlations in space (see Fig.~\ref{Fig:Intro} right). 
Instead of trying to identify avalanches we focus on this quantity and its correlation functions :
\begin{align}
C_v(x) &:= \langle v(0,t) \; v(x,t) \rangle - v_d^2 = v_d^2 \; \mathcal{F} \left( \frac{x}{\xi_v} \right) \, , \label{eq:Cv(x)} \\
G_v(\tau) &:= \langle v(x,t)\; v(x,t+\tau) \rangle - v_d^2 = v_d^2 \; \mathcal{G} \left( \frac{\tau}{t^*} \right)\, . \label{eq:Gv(tau)}
\end{align}
The proposed scaling forms rely on the existence of two scales : $\xi_v \sim v_d^{-\nu/\beta}$ and 
$t^* \sim v_d^{-\nu z/\beta}$.  
The first one is the correlation length at finite velocity and arises naturally from the combination of the scalings of the velocity $v\sim \left(f-f_c \right)^{\beta}$ and of the correlation length $\xi \sim (f-f_c)^{-\nu}$. 
The time scale $t^*$ is linked to $\xi_v$ through the dynamical exponent $z$ \cite{dynamical_exponent} : 
$t^* \sim \xi_v^z$.
Note that these assumptions are reasonable provided that $m^2$ is small enough, otherwise the parabolic potential confines the interface at length scales $\sim m^{-2/\alpha}$.

\smallskip
\textbf{Asymptotic forms}
We derive the asymptotic forms of $\mathcal{F}(y)$ and $\mathcal{G}(y)$ via a scaling analysis based on the existence of a unique correlation length (and a unique correlation time) when $v_d$ is small.
Below this length (and time), one expects to find the critical behavior while above it, the $f \to \infty$ behavior (equivalent to $v_d \to \infty$) should be recovered.
For a slow drive, $v_d \to 0$, the local velocity is intermittent: it takes values of order $v^{\max}$ (independent of $v_d$) with probability $\propto v_d$ and is almost zero otherwise.
The main contribution to $C_v(x)$ comes from the realizations for which both $v(0,t)$ and $v(x,t)$ are of order $v^{\max}$. 
In the critical regime, one expects from dimensional analysis that if $v(0,t)$ is of order $v^{\max}$, then $v(x,t)$ is also of order $v^{\max}$ with a probability that decays as $x^{-\beta/\nu}$~\cite{scaling_argument}. 
This gives $C_v(x) \sim v_d x^{-\beta/\nu} \sim v_d^2 \left( x/\xi_v \right)^{-\beta/\nu} $.
For temporal correlations, a similar reasoning yields 
$G_v(\tau) \sim v_d \tau^{-\beta/(\nu z)} \sim v_d^2 \left( \tau/t^* \right)^{-\beta/(\nu z)}$.

Concerning the large scale behavior, it is convenient to rewrite equation~\eqref{eq:Movement equation} in the comoving frame :
$u(x,t) \rightarrow v_d t + u(x,t)$
and neglect the parabolic drive.
The disorder becomes $\eta \left(x, v_d t + u(x,t) \right)$.
From dimensional analysis one sees that at large scales, when $x > \xi_v$ or $t > t^*$, $u$ is subdominant compared to $v_d t$ (see appendix B \cite{SupMat}).
Then the behavior of equation \eqref{eq:Movement equation} is captured by a linear Langevin equation that we solve in the appendix B \cite{SupMat}.
By plugging the solution into the correlation function we obtain~:
\begin{align}
C_v(x \gg \xi_v ) &\sim \left\lbrace 
\begin{array}{c}
1/x^{1+\alpha} \quad  \text{ for } \; \alpha < 2 \, ,\\
e^{-x/\xi_v} \quad \text{ for short-range ,}
\end{array} \right.\\
G_v(\tau \gg t^*) &\sim - \, 1/\tau^{1+\frac{1}{\alpha}} \quad \quad  \text{ for } \; \alpha \leqslant 2 \, .
\end{align}
\begin{flushleft}
\begin{figure}[t!]
\includegraphics[width=1\linewidth]{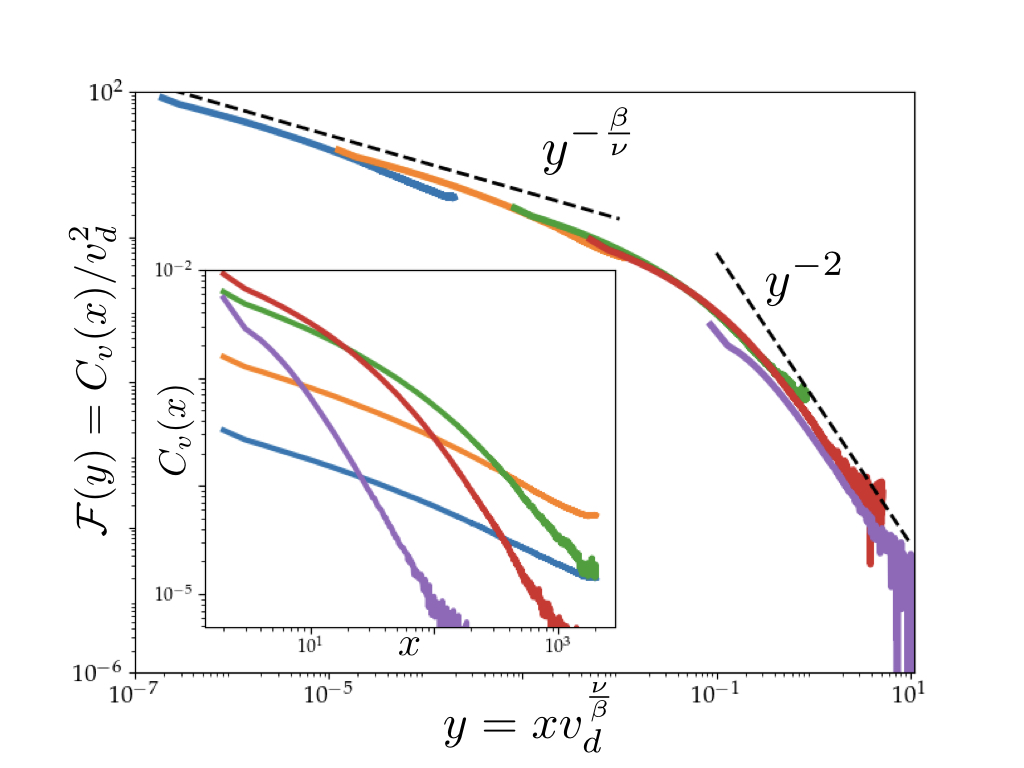}
\caption{Spatial correlations in the cellular automaton for driving velocities 
$v_d=0.002$ (blue), $0.01$ (yellow), $0.05$ (green), $0.1$ (red) and $0.3$ (purple). 
\textbf{Main panel :} A perfect collapse is observed using the scaling form \eqref{eq:Cv(x)}.
The asymptotic behaviors \eqref{eq:scaling form Cv(x)} are verified. In particular, going at large distances the decay $y^{-2}$ of the elastic interaction is recovered while at small distances the critical behavior 
$\beta/\nu \simeq 0.385$ is captured.
From the crossover the length $\xi_v$ is estimated to be $\xi_v \simeq 0.07 \, v_d^{-\nu/\beta}$.
\textbf{Inset :} Nonrescaled correlation function $C_v(x)$.
System size : $L = 4096$, mass: $m^2 = 10^{-3}$}
\label{Fig:Cv(x)_simus}
\end{figure}
\end{flushleft}
Note that at large distance, the decay of the spatial correlation function provides exactly the range $1 + \alpha$ of the elastic interactions.
Interestingly, the long time behavior of $G_v(\tau)$ displays anticorrelations with an $\alpha$ dependent power law decay.
We note a qualitative similarity with the anticorrelation between the sizes of dynamical avalanches, predicted and numerically measured in \cite{ledoussal2019}.
Collecting all these informations, we can write the full scaling forms (here for $\alpha=1$, i.e. for crack and wetting fronts) :
\begin{align}
\mathcal{F}(y) &\sim \left\lbrace 
\begin{array}{c}
y^{-\frac{\beta}{\nu}} \quad  \text{ if } \; y \ll 1 \, , \\
y^{-2} \quad \text{ if } \; y \gg 1 \, ,
\end{array} \right. \label{eq:scaling form Cv(x)} \\
\mathcal{G}(y) &\sim \left\lbrace 
\begin{array}{c}
y^{-\frac{\beta}{\nu z}} \quad \text{ if } \; y \ll 1 \, , \\
-y^{-2} \quad \text{ if } \; y \gg 1 \, .
\end{array} \right. \label{eq:scaling form Gv(tau)}
\end{align}

\smallskip
\textbf{Simulation and experiment}
We implemented a cellular automaton version of the variant of equation \eqref{eq:Movement equation} with $f$ replaced by $m^2 \left( v_d t - u(x,t) \right)$ and $\alpha=1$. The three variables $u$, $x$ and $t$ are integer.
In particular we assume periodic boundary conditions along $x$ which takes values ranging from $0$ to $L-1$.
The local velocity is defined as :
\begin{align}
v(x,t) &= \theta \Large( F(x,t) +  \eta\left(x, u(x,t)\right) \Large) \, , \\
F(x,t) &= m^2 \left( v_d t - u(x,t) \right) + \sum_{x'} \frac{u(x',t) - u(x,t)}{|x'-x|^{2}} \, , \notag
\end{align}
$\theta$ being the Heaviside function.
Here the quenched disorder  pinning force $\eta$ should be negative and uncorrelated. In practice, we take identical and independent variables whose distribution is the negative part of the normal law. At each time step all the points feeling a positive total force jump one step forward while the other points - which feel a negative force - stay pinned.
Then the time is incremented, $t\rightarrow t+1$, and the forces are recomputed : new pinning forces are drawn for the jumping points, the elastic force is updated by using a Fast Fourier Transform (FFT) algorithm and the driving force is incremented by $m^2 v_d$.
For the numerical implementation, we started from a flat configuration, turned on the dynamics and waited until reaching the steady state before computing the two correlation functions.

\begin{figure}[t!]
\includegraphics[width=1\linewidth]{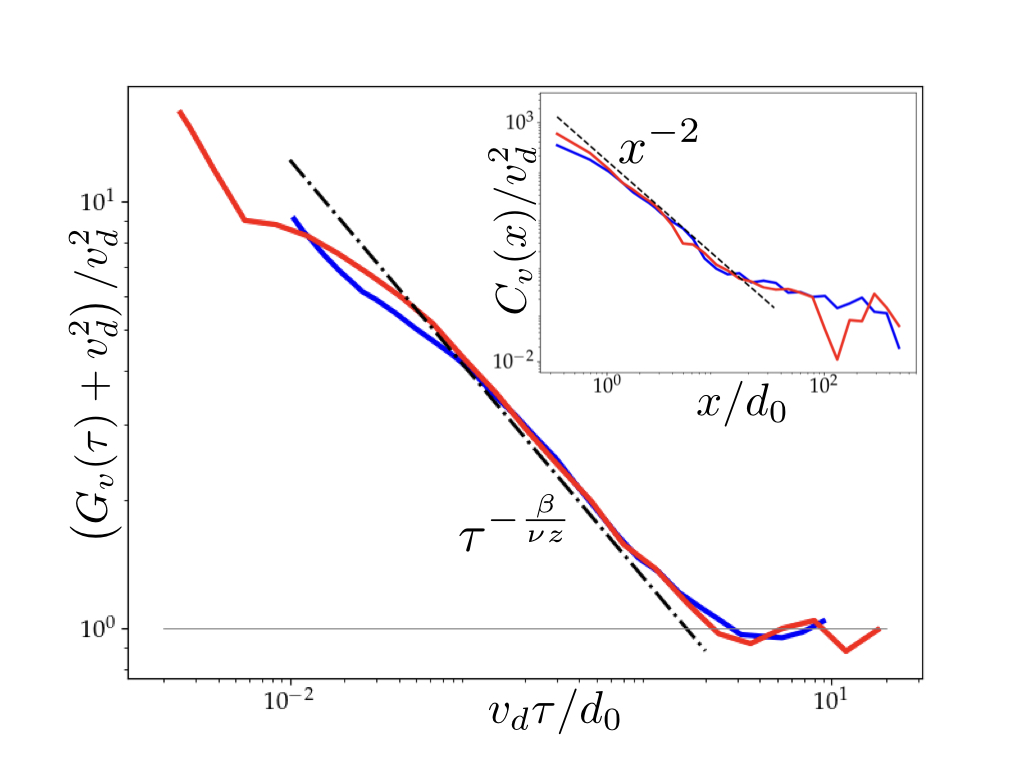}
\caption{\textbf{Inset :} Spatial correlations in the experiment for driving velocities 
$v_1 = 132$ nm/s (blue) and $v_2=31$ nm/s (red). The large distance decay $x^{-2}$ of the elastic interactions is observed. No rescaling of the x-axis was performed. 
\textbf{Main panel :} Temporal correlations for the same velocities.
The asymptotic predictions of~\eqref{eq:scaling form Gv(tau)} are verified :
anticorrelations are observed at large time and the depinning power law decay is recovered at short time.
The time axis was rescaled by $v_d$.
\label{Fig:Correlations_manip}}
\end{figure}

The experimental data presented here correspond to planar crack propagation. 
A $5 \, \mathrm{mm}$ thick plexiglas plate is detached from a thick silicone substrate using the beam cantilever geometry depicted in Fig.~\ref{Fig:Intro}~\cite{chopin2018}. To introduce disorder, we print obstacles of diameter $d_0 = 100~\mu\mathrm{m}$ with a density of $20\,\%$ on a commercial transparency that is then bonded to the plexiglas plate. 
Crack front pinning results from the strong adhesion of the ink dots to the substrate. Images of $1800 \times 1800$ pixels are taken normal to the mean fracture plane every second. As the system is fully transparent, the crack front appears as the interface between the clear and the dark region observed on the image.
The pixel size is $35~\mu\mathrm{m}$, so the observed front length is $63~\mathrm{mm}$.

We tested two different velocity regimes : $v_1= 132 \pm 3~\mathrm{nm.s}^{-1}$ and $v_2 = 31 \pm 1~\mathrm{nm.s}^{-1}$. 
The local crack speed is computed using the methodology proposed in Refs.~\cite{maloy2006,tallakstad2011} 
based on the waiting time matrix : the number of frames during which the front stays inside each pixel provides the waiting time in this pixel, from which the local speed is inferred. 

\begin{figure}[t]
\includegraphics[width=1\linewidth]{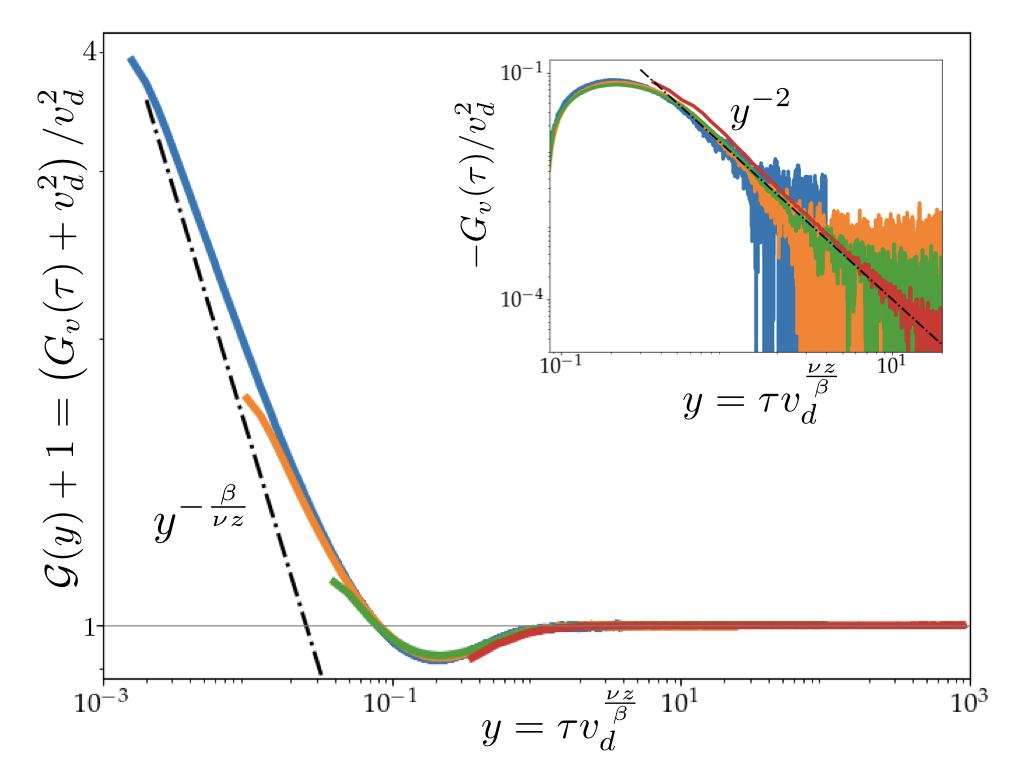}
\caption{Temporal correlations in the cellular automaton for driving velocities $v_d=0.02$ (blue), $0.05$ (yellow), $0.1$ (green) and $0.3$ (red). 
\textbf{Main panel :} The scaling form \eqref{eq:Gv(tau)} and the asymptotic behaviors \eqref{eq:scaling form Gv(tau)} are verified.
In particular at large time we observe anticorrelations
while at small time the depinning behaviour $y^{-\beta/(\nu z)}$ is recovered ($\beta/(\nu z) \simeq 0.500$).
\textbf{Inset :} Zoom on the anticorrelation. For $\alpha=1$ the expected decay is $1/y^2$.
System size : $L$ and $m$ range from $L=2048$, $m^2=10^{-3}$ ($v_d=0.3$) to $L=32768$, $m^2=10^{-4}$ ($v_d=0.02$).
\label{Fig:G(tau)_simus}}
\end{figure}

Both experiments and the cellular automaton are expected to belong to the universality class of a one dimensional interface with $\alpha=1$. The depinning exponents of this class have been computed numerically~: $\zeta=0.388 \pm 0.002$ \cite{rosso2002}, $\nu= 1/(1-\zeta) = 1.634 \pm 0.005$, $\beta = 0.625 \pm 0.005$, 
$z=0.770 \pm 0.005$ \cite{duemmer2007} in agreement with renormalization group calculations \cite{ledoussal2002}.
The spatial correlations of the local velocity are shown on Figs.~\ref{Fig:Cv(x)_simus} and \ref{Fig:Correlations_manip}. The results of the simulation perfectly collapse on the scaling form \eqref{eq:Cv(x)} showing that a unique correlation length $\xi_v$ controls the dynamics. The asymptotic form proposed in \eqref{eq:scaling form Cv(x)} is verified, in particular the decay in $1/x^2$ is the fingerprint of the long-range nature of the elasticity.

Our experiment confirms the large distance decay as $1/x^2$. This proves that the elastic kernel of the crack front is long-range in this experiment~\cite{chopin2018}. 
For both velocities the large scale behavior breaks down for distances of $2$-$3$ pixels.
This is consistent with our estimation $\xi_v \simeq 2 d_0$ at the end of appendix A \citep{SupMat}.
At variance with the simulation,
varying the crack speed $v_d$ does not affect the scale $\xi_v$.
This rather counter-intuitive behavior results from the velocity dependence of the material toughness \cite{kolvin2015}.
This induces that the characteristic mobility $\mu$ involved in equation~\eqref{eq:Movement equation} scales with the mean crack speed $v_d$ so that the distance $v_d/\mu$ to the critical point remains constant \cite{chauve2000}
(see \cite{chopin2018} and the last section of appendix A \cite{SupMat}).

We now turn to the temporal correlation function. The results are shown on Figs.~\ref{Fig:Correlations_manip} and \ref{Fig:G(tau)_simus} where we plot the correlation
function $G_v(\tau) + v_d^2$ and normalize it by dividing by $v_d^2$.
The parts of the curves below $1$ correspond to anticorrelation. Again numerical simulations show a perfect collapse on the scaling form \eqref{eq:Gv(tau)} with a unique $t^*$ and the 
asymptotic form of equation~\eqref{eq:scaling form Gv(tau)} is verified :
the anticorrelation displays a power law decay $1/\tau^{2}$ (see inset in Fig.~\ref{Fig:G(tau)_simus})
and the exponent $\beta/(\nu z) \simeq 0.50$ at small scale is recovered.
It is remarkable that the power law behavior $y^{-\beta/\nu z}$ holds for the non connected function $\mathcal{G}(y)+1$ until the time when anticorrelation appears.
A similar behavior with a crossover from a power law decay to anticorrelation is observed in the experiment.
However, curves corresponding to different crack speeds are collapsed using $v_d$ instead of $t^* = v_d^{\beta/(\nu z)}$. This is also explained by the relation $\mu \sim v_d$ specific to our material.
This is the first time that anticorrelation is predicted and observed in depinning systems at finite drive (see also appendix C \cite{SupMat}).
At short time the scaling behavior $G_v(\tau) \sim \tau^{-\beta/\nu z}$ holds when $\xi_v$ is large compared to the microscopic scale of the disorder. 
Otherwise a crossover to a different regime, not studied here, should occur at very small scales. 
The power law decay observed here is consistent with the depinning prediction $\tau^{-\beta/\nu z}$
even if $\xi_v$ is of the order of $d_0$.

\smallskip
\textbf{Discussion}
Our findings open new perspectives for the experimental study of disordered elastic interfaces. As the correlations of the local velocity display universal features of the depinning even when the driving speed is finite, the critical behavior can be investigated far from the critical point.  
This provides a robust and efficient method to identify the universality class of the transition and to test the relevance of specific depinning models. 

The analysis of the local speed correlations has already been performed in previous simulations and experiments. But the link with the critical exponents was missing. In the simulations of Ref.~\cite{duemmer2005} of an interface with short-range elasticity, the correlation function $C_v(x)$ was used to extract the scale $\xi_v$ and the exponential cutoff was observed but the small scale exponent $\beta/\nu$ was not predicted. In the fracture experiments of Tallakstad {\it et al.}~\cite{tallakstad2011}, the correlation functions of the local velocity were found to scale as $C_v(x) \sim x^{-\tau_x}$ and $G_v(\tau) \sim \tau^{-\tau_t}$
with exponents $\tau_x = 0.53 \pm 0.12$ and $\tau_t = 0.43$ a bit away from the depinning 
predictions $\beta / \nu \simeq 0.38$ and $\beta / (\nu z) \simeq 0.50$. 
However exponential cutoffs  at large distances and time were used for the fit and the anticorrelation in time was not observed.
Note that standard log-log plot routines discard negative values and one must use alternative plots to see the anticorrelation. It would be interesting to test how far  the behavior predicted in this study could capture the Tallakstad {\it et al.}'s experiments, as their systems allow the exploration of the crack behavior closer to the critical point than the one used in this study.
Finally we note that Gjerden \textit{et al}.~\cite{gjerden2014} computed the same correlation functions in simulations
of a fiber bundle model that mimics the presence of damages in front of the crack.
Their model should fall into the depinning universality class with long-range
elasticity~\cite{gjerden2014} and they measured $\tau_x=\tau_t = 0.43$ with cutoffs faster than exponential.
 
Finally it is important to remark that the scaling forms~\eqref{eq:Cv(x)} and \eqref{eq:Gv(tau)} are very general and valid for all out-of-equilibrium transitions with avalanche dynamics. The asymptotic forms~\eqref{eq:scaling form Cv(x)} and \eqref{eq:scaling form Gv(tau)} are also very general, as beyond $\xi_v$ 
the spatial correlations decay as $1/x^{d+\alpha}$ for a long-range model ($d$ being the spatial dimension) and 
exponentially fast for short-range elasticity. It would certainly be insightful to probe this behavior in various problems, including those where the nature of the elastic interactions still needs to be deciphered 
or in the context of the yielding transition where avalanches of plastic events are observed \cite{lin2014}.

\begin{acknowledgments}
{\it Acknowledgments:}
We thank E. Bouchaud and V. D{\'e}mery for useful discussions.
\end{acknowledgments}

\bibliography{BiblioScaling}

\begin{thebibliography}{43}
\bibitem[{sca()}]{scaling_argument}
\bibinfo{note}{Indeed as $ \langle v(x,t)|v(0,t) \sim v_{max} \rangle = v_d
  \mathcal{H}(x/\xi_v )$ in order to have a finite result when $ v_d\to 0 $ one
  must have $ \mathcal{H}(z) \sim z^{-\beta/\nu}$ when $z \to 0$}.
\end{thebibliography}

\appendix

\newpage

\begin{widetext}

\bigskip

\bigskip

\begin{large}
\begin{center}

Supplemental Material for {\it Universal scaling of the velocity field in crack front propagation}

\end{center}
\end{large}

We give the details of some of the calculations described in the main text of the Letter.
In appendix A we derive the equation of motion for the crack front. This equation is discussed in the literature but we rederive it in order to be self-contained and accessible to a general audience. In the last section of the appendix we modify the equation to accounts for the visco-elasticity of  the silicone substrate used in the experiment. This modification explains why  the change of $v_d$ of a factor $4$  does not affect the length $\xi_v$.
In appendix B we compute explicitly the correlation functions in the thermal limit which correspond to the large scales asymptotic behaviors provided in equations (5) and (6) of the main text.
Finally in appendix C we give more evidence about the anticorrelation observed in the experiment at large time.

\section{A) Equation of motion for  crack propagation in disordered materials}\label{Appendix A}

In our experiment a crack propagates at constant velocity $v_d$. The derivation of the equation of motion for the  front in presence of impurities can be found in \cite{chopin2018}. 
Here for the sake of completness we recall the main steps of the derivation and provide an explanation for the surprising observation that the data with $v_d= 132 \pm 3 \ nm.s^{-1}$ and $v_d= 31 \pm 1 \ nm.s^{-1}$ seem to display the same distance from the critical depinning point.

\subsection{Crack propagation in homogeneous elastic material}

\begin{figure}[b]
\includegraphics[width=.5\linewidth]{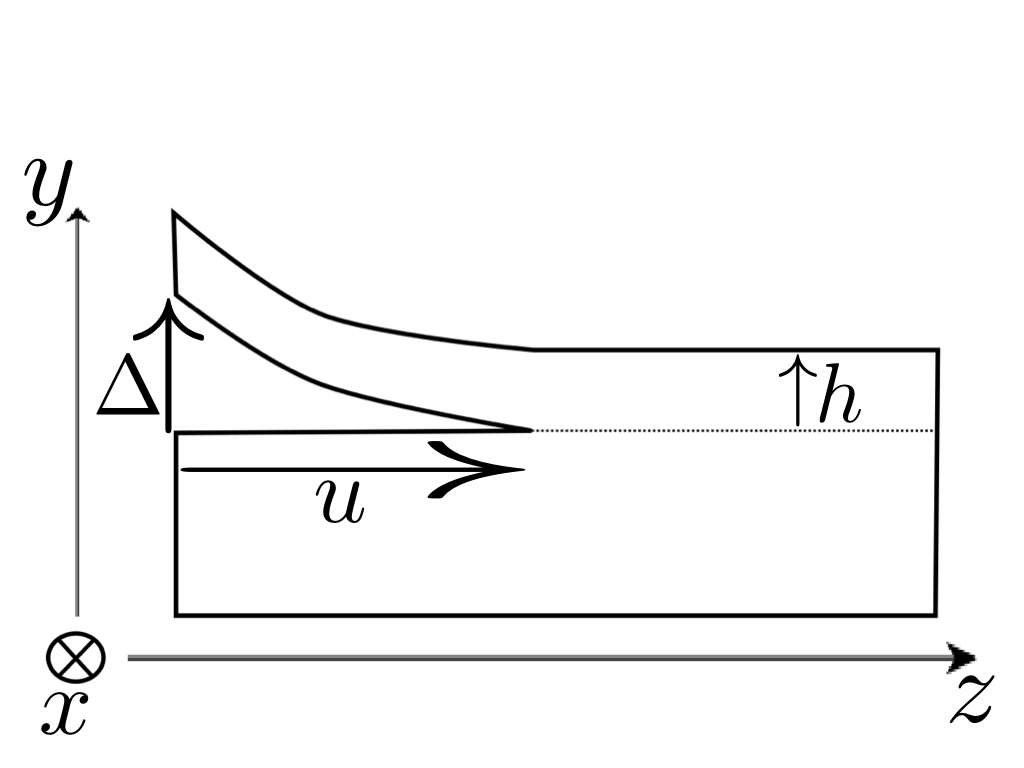}
\caption{Sketch of a crack of length $u$ for an homogeneous elastic material of unit width (along the $x$ direction).
The upper plate has height $h$ and a vertical displacement $\Delta$ is imposed at its end. This plate can be described 
as an Euler-Bernoulli cantilever beam.
\label{Fig:Sketch crack}}
\end{figure}

It is convenient to start with the homogeneous elastic material. Here the front is prefectly flat and is characterized by its position $u$ and  speed $\dot u$ (see Fig. \ref{Fig:Sketch crack}).
Note that a priori the crack front position is a vector $\vec u = (u_y, u_z)$. However in our experiment the crack propagates at the interface between two materials hence it is natural to assume in-plane propagation where $u_y$ stays constant and only $u_z:=u$ evolves. 
In a real homogeneous material one should follow the in-plane and out-of-plane propagation of $\vec u$ as done in Ref.~\cite{basu2019} where however the long-range elasticity of equation (1) of the main text has been replaced by short-range elasticity.
Following  Griffith's idea \cite{griffith1920},
the  evolution in time of the crack  is determined by the energy balance (per unit surface) between the energy released when the material is fractured and the energy needed to create new fracture surfaces :
 \begin{equation}\label{eq:Griffith criterion}
G^{\text{dyn}}(u, \dot u) = G_c \, .
\end{equation}
$G_c$ is the fracture energy, which is constant for homogeneous elastic materials. 
$G^{\text{dyn}}$ is the energy release rate which accounts for the release of potential elastic energy minus the kinetic term.
It displays a simple velocity dependence~\cite{freund1990}~:
\begin{equation}\label{eq:Freund relation}
G^{\text{dyn}}(u, \dot u) = \left( 1- \frac{\dot u}{c_R} \right) G^{\text{el}}(u) \, ,
\end{equation} 
where $c_R$ is the Rayleigh wave velocity and $G^{\text{el}}(u)$ is the elastic energy release rate. 

In the experimental setup sketched in Fig. \ref{Fig:Sketch crack} we impose a displacement $\Delta$ at the end of the upper plate. The elastic energy associated with the deformation of the plate is a function of the imposed
displacement $\Delta$ and of the crack length $u$. 
In particular if one describes the plate as an Euler-Bernoulli cantilever beam of unit width and height $h$, the 
elastic energy writes~\cite{freund1990}~:
\begin{equation}\label{eq:Beam elastic energy}
E^{\text{el}}(u, \Delta) = \frac{E h^3 \Delta ^2}{8 u^3} \, ,
\end{equation}
with $E$ the Young modulus. 
The elastic energy release rate is then 
\begin{equation}
G^{\text{el}} (u, \Delta) = - \frac{d E^{\text{el}}}{du} (u, \Delta) = \frac{3}{8} \frac{E h^3 \Delta ^2}{u^4} \, .
\end{equation}

The experiment starts by imposing an initial displacement $\Delta_0$ which opens the crack up to a length $u_0$ 
such that $G^{\text{el}} (u_0, \Delta_0) = G_c$.
Then the displacement is increased as $\Delta(t) = \Delta_0+v_0 t$ and 
the crack  moves from $u_0$ to $u_0 + u(t)$. Keeping  $v_0 t \ll \Delta_0$, $u(t) \ll u_0$ one can write the first order expansion of the elastic energy release rate :
\begin{equation}\label{eq:expansion loading}
 G^{\text{el}}(u_0 + u(t), \Delta_0 + v_0 t) = G_c   +  \dot G_0 t + G_0' u(t) \, ,
\end{equation} 
where $\dot G_0 = v_0 \partial_{\Delta} G^{\text{el}}(u_0,\Delta_0)$ and 
$G_0' = \partial_u G^{\text{el}}(u_0,\Delta_0)$.
When $\dot u \ll c_R$, combining equation \eqref{eq:expansion loading} with equations \eqref{eq:Griffith criterion} and \eqref{eq:Freund relation} to first order yields the following equation of motion :
\begin{equation}
\frac{1}{\mu} \dot u =  k \,\left( v_d t - u \right) \, , 
\end{equation}
with $\mu = c_R$, $v_d=-\frac{\dot G_0}{ G'_0} = \frac{u_0}{2\Delta_0} v_0$ and $k= -\frac{G'_0}{G_c} = \frac{4}{u_0}$.
Thus by varying $v_0$ one can control the steady velocity $v_d$ of the crack propagation.

\subsection{Crack propagation in disordered elastic material}

When the material is heterogeneous the fracture energy displays local fluctuations around its mean value $G_c$ :
\begin{equation}\label{eq:local fluctuations}
G_c(x,u) = G _c + \delta G_c (x, u) \, .
\end{equation}
As a consequence the crack front $u(x,t)$ becomes rough. This non trivial shape  introduces a correction in the elastic energy release rate, which was computed to first order in perturbation by Rice~\cite{rice1985}~:
 \begin{equation}\label{eq:Rice1985}
G^{\text{el}}\left(x, u(x,t), \Delta \right) = G^{\text{el}}\left(\overline{u}(t), \Delta\right)  \left(1 + \frac{1}{\pi} \int \frac{u(x',t)-u(x,t)}{|x'-x|^2}dx' \right)
\end{equation}
where $\overline{u}(t)= L^{-1}\int u(x,t) dx$ is the average front position. 
The balance between the energy release and the fracture energy still holds but must now be written at the local level :
\begin{equation}\label{eq:Griffith criterion_local}
G^{\text{dyn}}(x,u(x,t), \dot u(x,t)) =  \left( 1- \frac{\dot u(x,t)}{c_R} \right) 
		G^{\text{el}}\left(x, u(x,t), \Delta \right) = G_c(x,u(x,t)) \, .
\end{equation}
In presence of impurities the first order expansion of $G^{\text{el}}$ becomes :
\begin{equation}\label{eq:expansion loading2}
G^{\text{el}}(x, u(x,t),t)=   G_c   +  \frac{G_c}{\pi} \int \frac{u(x',t)-u(x,t)}{|x'-x|^2}dx' + \dot G_0 t+G_0'  \overline{u}(t) \, .
\end{equation} 
By combining together equations \eqref{eq:expansion loading2}, \eqref{eq:Griffith criterion_local}, \eqref{eq:local fluctuations} and \eqref{eq:Freund relation} one obtains :

\begin{equation}\label{eq:Movement equation appendix}
\frac{1}{\mu} \dot u =  k \,\left( v_d t - \overline{u}(t) \right) + \frac{1}{\pi} \int \frac{u(x',t)-u(x,t)}{|x'-x|^2}dx'  - \frac{\delta G_c}{G _c} \, ,
\end{equation}
where $\mu=c_R$.
This equation is equivalent to equation (1) of the main text :
$k$ plays the role of $m^2$,
$- \frac{\delta G_c(x,u)}{G _c}$ is the disorder and the elasticity is long-range. However in equation (1)
$\overline{u}(t)$ has been replaced by $u(x,t)$.
Note that $k=\frac{4}{u_0}$ hence if $u_0$ is large enough the length $k^{-1}$ is much larger than $\xi_v$.
In our experiment $u_0$ is of the order of a few centimeters which is much larger than the fluctuations of the front. This justifies the small mass assumption in the main text.

\subsection{Crack propagation law in visco-elastic materials}
In our experiment we used a plexiglas plate of PMMA (polymethyl methacrylate) glued on a silicone substrate of PDMS (polydimethylsiloxane).
The PDMS is not perfectly elastic but displays a visco-elastic behaviour. This impacts the fracture energy $G_\mathrm{c}$ that shows a rather strong dependence with crack speed~\cite{chopin2018, kolvin2015}~: 
\begin{equation}
G_\mathrm{c} = G_\mathrm{c}(\du) \simeq \left( 1 + \frac{\du}{v_c} \right)^{\gamma} \, .
\end{equation} 
In particular for our experiment, we have $v_c \ll v_2$ and $\gamma \simeq 1/3$~\cite{chopin2018}, 
so that $G_c(\du) \sim \du ^{\gamma}$ in the range of crack speeds investigated. Hence the expansion for the fracture energy in presence of impurities should be modified as follow :
\begin{equation}\label{eq:G_c velocity expansion}
\frac{G_c(x,u,\dot u)}{G _c(v_d)} = 1 + \frac{\delta G_c(x,u, v_d)}{G _c(v_d)} + \frac{\gamma}{v_d} (\dot u-v_d) \, .
\end{equation}
The last term in \eqref{eq:G_c velocity expansion} modifies the equation of motion \eqref{eq:Movement equation appendix} as follow:
\begin{equation}\label{eq:Movement equation appendix 2}
\left( \frac{1}{c_R} + \frac{\gamma}{v_d} \right) v = k \,\left( v_d t - \overline{u}(t) \right) - \frac{\delta G_c}{G_c} + \frac{1}{\pi} \int \frac{u(x')-u(x)}{|x'-x|^2}dx' 
\end{equation}
where the constant $\gamma$ has been absorbed in the loading. 
Note that equations \eqref{eq:Movement equation appendix} and \eqref{eq:Movement equation appendix 2} have the same
form but the mobility has been renormalized. In the experiment the driving velocity satisfies $v_c \ll v_d \ll c_R$ so that $ \mu \simeq \frac{v_d}{\gamma}$.
Functional Renormalization Group calculations have shown that the dynamical correlation length $\xi_v$ depends not only on the driving velocity $v_d$ but also on the mobility~cite{chauve2000}~: 
\begin{equation}
\xi_{v} = L_c \left( \frac{\mu f_c}{ v_d} \right)^{\frac{\nu}{\beta}}
\end{equation}
where $L_c$ is the Larkin length and $f_c$ the critical force. For our experimental conditions the ratio 
$\frac{\mu}{v_d}$ does not depend on $v_d$. Hence tuning $v_d$ does not change $\xi_v$ and we cannot come closer to the 
depinning critical point.
$L_c$ and $f_c$ have been estimated to be $L_c = d_0/\sigma^2$ and $f_c = \sigma^2$ \cite{demery2014}
where $\sigma^2 = \left\langle \delta G_c^2 \right\rangle$.
In our system the disorder is controlled and we have $\sigma \simeq 1/2$. A numerical application hence yields $\xi_v \simeq 2 d_0$.

\section{B) Thermal approximation for the large scale behavior of the correlation functions}

For the purpose of computing the tails of the correlation functions 
it is convenient to rewrite equation (1) of the main text in the 
comoving frame : $u(x,t) \rightarrow v_d t + u(x,t)$. The disorder term then becomes $\eta\left(x, v_d t + u(x,t) \right)$.
To describes large scales $x > \xi_v$ or $t > t^*_v$, a reasonable approximation is to use an effective model
where the disorder is replaced by white noise. 
Its correlations then read :
\begin{equation}\label{eq:Correlations disorder}
\langle \eta\left(0, u(0,0)\right) \eta\left(x, v_d \tau + u(x,\tau)\right)\rangle = \delta(x)\delta\left(v_d\tau +\Delta u (x,\tau)\right) \, 
\end{equation}
where $\Delta u (x,\tau) := u(x,\tau) - u(0,0)$.
From dimensional analysis $\tau \sim x^z$ while $\Delta u \sim x^{\zeta}$. 
Hence $v_d \tau / \Delta u \sim \xi_v ^{-\beta / \nu} x^{z-\zeta} \sim \left(x/\xi_v \right)^{\beta/\nu}
\sim \left(\tau/t^* \right)^{\beta/(\nu z)}$.
So we see that at large length scale, $x > \xi_v$ or large time scale, $\tau > t^*$, $\Delta u (x,\tau)$ is subdominant compared to $v_d \, \tau$. The disorder can thus be replaced on these scales
by an effective thermal noise $\eta (x, v_d t)$ and 
the equation of motion becomes a Langevin equation~:
\begin{align}
\partial_t u(x,t) &= -m^{2}u(x,t) + \eta\left(x, v_d t \right) + \int \frac{u(x',t) - u(x,t)}{|x'-x|^{1+\alpha}} dx' \, ,  \label{eq: Langevin eq. real space} \\
\partial_t u(q,t) &= -m^{2}u(q,t) + \eta\left(q, v_d t \right) -|q|^{\alpha} u(q,t) 
 		= -b(q)u(q,t) + \eta\left(q, v_d t \right) \, . \label{eq: Langevin eq. fourier space}
\end{align}
where $b(q) = m^2 + |q|^{\alpha}$.
The Fourier transformation from equation \eqref{eq: Langevin eq. real space} to \eqref{eq: Langevin eq. fourier space} holds for $0 < \alpha < 2$. 
Equation \eqref{eq: Langevin eq. fourier space} with $\alpha = 2$ corresponds to the short-range (SR) elasticity.
It is a first order linear differential equation and its solution reads :
\begin{equation}\label{eq:solution Langevin equation}
u(q, t) = \int_{-\infty}^{t} e^{-b(q) (t-t')}  \eta(q,v_d t') dt' \, .
\end{equation}

The general two-point correlation function, which we denote $C(x,\tau)$ can be written as a Fourier transform~:
\begin{equation}\label{eq:General 2-points correlations}
C(x,\tau) := \langle \partial_t u(y,t) \partial_t u(y+x,t + \tau) \rangle = \int \frac{dq_1 dq_2}{(2\pi)^2} e^{iq_1 y} e^{iq_2(x+y)}
			 \langle \partial_t u(q_1,t) \partial_t u(q_2,t + \tau) \rangle \, .
\end{equation}
We can compute the correlation term in Fourier space by plugging in $\partial_t u(q,t) = \eta(q,v_d t) - b(q) \int_{-\infty}^{t} e^{-b(q) (t-t')}  \eta(q,v_d t') dt'$ with It\^o convention and using the noise correlation 
$\langle \eta(q_1,v_d t_1) \eta(q_2,v_d t_2) \rangle =  2 \pi \delta(q_1+q_2) \delta(t_1-t_2) / v_d$ :
\begin{align}
\langle \partial_t u(q_1,t) \partial_t u(q_2,t + \tau) \rangle
	&= \frac{ 2 \pi \delta(q_1+q_2)}{v_d} \left( \delta(\tau) - b(q_2) e^{-b(q_2)\tau} + b(q_1)b(q_2)
	\frac{e^{-b(q_2)\tau}}{b(q_1)+b(q_2)} \right) \, .
\end{align}
Performing one integral over $q$, equation \eqref{eq:General 2-points correlations} now reads :
\begin{equation}\label{eq:General 2-points correlations simplified}
C(x,\tau) =  \frac{1}{v_d} \delta(x) \delta(\tau) 
		- \frac{1}{2 v_d} \int \frac{dq}{2\pi} e^{iqx} b(q) e^{-b(q)\tau} \, .
\end{equation}
The first term is local and originates from the delta function approximation to the noise,
and represents the correlations at shorter scales, not described accurately by the present effective model.
We now focus on the second term which describes the large scale tail.

Let us indicate the result for $\alpha=1$, our case of most interest. Let us set the mass to zero. One finds,
at large scales $x > \xi_v$ or $\tau > t^*_v$,
\begin{equation}\label{alpha1} 
C(x,\tau) \simeq  \frac{1}{2 v_d} \partial_\tau \int \frac{dq}{2\pi} e^{iqx} e^{-|q| \tau} 
= \frac{1}{2 \pi v_d} \partial_\tau \frac{\tau}{x^2 + \tau^2} = \frac{1}{2 \pi v_d}  \frac{x^2-\tau^2}{(x^2 + \tau^2)^2}
\end{equation}

Consider now the spatial correlations, setting $\tau=0$. For $\alpha=1$ we find the decay
\begin{equation}
C_v(x) 
	    \underrel{x\gg\xi_v}{\simeq} \frac{1}{2 \pi v_d}  \frac{1}{x^2} 
\end{equation}
This result extends to general $0<\alpha \leq 2$ as follows :
\begin{equation}\label{eq:Cv(x)_appendix}
C_v(x) \underrel{x\gg\xi_v}{\simeq} \, \frac{\Gamma(1+\alpha) \sin \left(\frac{\pi \alpha}{2} \right)}{2\pi v_d} \,  \frac{1}{x^{1+\alpha}} \quad \text{for} \quad \alpha \leqslant 2 \, .
\end{equation}
It can be obtained, e.g. by introducing a regularization factor $e^{-\epsilon |q|}$ 
in the inverse Fourier transform of $|q|^{\alpha}$ and 
taking the limit $\epsilon \to 0$ at the end, using:
\begin{align}
\int \frac{dq}{2\pi} e^{iqx-\epsilon |q|} |q|^{\alpha} &= \frac{\Gamma(1+\alpha)}{\pi (\epsilon^2+x^2)^{\frac{1+\alpha}{2}}}
   \cos \left( (1+\alpha) \arctan \left(\frac{x}{\epsilon} \right) \right) \, .
\end{align}

For the short-range elasticity ($\alpha=2$) the prefactor in \eqref{eq:Cv(x)_appendix}
vanishes, and the large distance decay is no more a power law within this model,
but is much faster. Although a quantitative treatment goes beyond the present
effective model, one can obtain some qualitative idea by considering 
a model with a finite correlation length along $x$, e.g. replacing $\delta(x)\rightarrow e^{-\frac{|x|}{\ell}}$.
The disorder correlator then becomes :
\begin{equation}
\langle \eta(q_1, v_d t_1) \eta(q_2, v_d t_2) \rangle = \frac{2 \ell}{1+(q_2 \ell)^2} \, (2 \pi) \delta(q_1+q_2) \, \frac{\delta(t_1-t_2)}{v_d} \, . 
\end{equation}
When computing the spatial correlation function in the limit $m\to 0$, we obtain, discarding all $\delta(\tau)$ and
$\delta(x)$ terms (i.e. assuming $\ell$ is the largest length)
\begin{equation}
C_v(x) \underrel{x\gg\ell}{\simeq} - \frac{1}{v_d \ell} \int \frac{dq}{2 \pi} e^{i q x} \frac{(\ell q)^2}{1+ (\ell q)^2}
\simeq \frac{1}{2 v_d \ell^2} e^{- |x|/\ell} 
\end{equation}
which is the rationale for the exponential decay quoted in the main text (6). It is then
reasonable to expect that $\ell$ will be of order $\xi_v$. Again, this is not at the
present stage an accurate calculation which would require to account for more details about
the renormalized disorder.

We now turn to the temporal correlations. Consider first $\alpha=1$. We obtain,
setting $x=0$ and $\tau > 0$ in \eqref{alpha1} 
\begin{equation}
G_v(\tau) \underrel{\tau \gg t^*}{\simeq}  - \frac{1}{2 \pi v_d}  \frac{1}{\tau^2} \, .
\end{equation}
For general $\alpha$ we obtain
\begin{align}
G_v(\tau) &\underrel{\tau \gg t^*}{\simeq} -\, \frac{1}{2v_d} \int \frac{dq}{2\pi} b(q)e^{-b(q)\tau} 
\underrel{\tau \gg t^*}{\simeq} -\, \frac{e^{-m^2\tau}}{2\pi v_d} \int_0^{\infty}  dq (m^2 + |q|^{\alpha}) e^{-|q|^{\alpha} \tau} \underrel{\tau \gg t^*}{\simeq} -\, \frac{e^{-m^2\tau} (1+m^2\alpha\tau)\Gamma\left(\frac{1}{\alpha}\right)}{2\pi v_d \alpha^2} \; \frac{1}{\tau^{1+\frac{1}{\alpha}}} \, .
\end{align}
In the limit $m \to 0$ this reduces to :
\begin{equation}\label{eq:Gv(tau)_appendix_thermal}
G_v(\tau) \underrel{\tau \gg t^*}{\simeq} -\,\frac{\Gamma\left(\frac{1}{\alpha}\right)}{2\pi v_d \alpha^2} \; \frac{1}{\tau^{1+\frac{1}{\alpha}}} \, .
\end{equation}
Several remarks are in order. First we note that, at variance with the spatial decay,
the temporal decay remains a power law even for local elasticity, a property of standard diffusion
itself. Second, the negative sign in front of the result is the mark of anticorrelations. 
Although the regime described here $\tau > t^*_v$ is far from the intermittent one $\tau < t_v^*$, this is
in qualitative agreement with the anti-correlation of dynamical avalanches 
found in \cite{ledoussal2019} 
(see in particular the Fig. 5 there). We can thus
expect a robust region of negative temporal correlations in a broad region of time scales,
as observed. 
Note that in \eqref{alpha1} the spatio-temporal correlation changes sign along the
line $x=\tau$ (in the present units where all elastic and dynamic coefficients have been set to unity),
which would be nice to observe. 
Equations \eqref{eq:Cv(x)_appendix} and 
\eqref{eq:Gv(tau)_appendix_thermal} correspond to equations (6) and (7) of the main text.

Note that the approach used in this section, based on replacing the quenched noise with a velocity dependent thermal noise, does not allow to recover the correct dependence on $v_d$ but only the large scale 
dependence on $\tau$ and $x$. Indeed, in the replacement \eqref{eq:Correlations disorder}
we have not tried to be accurate: one could refine the model by multiplying 
by a prefactor with the correct dimension, and appropriate dependence in velocity (which 
could in principle be predicted by the renormalization group \cite{chauve2000}
which goes beyond this study).

\section{C) Statistical significance of the anticorrelation observed in the experiment}

\begin{figure}[b]
\includegraphics[width=.5\linewidth]{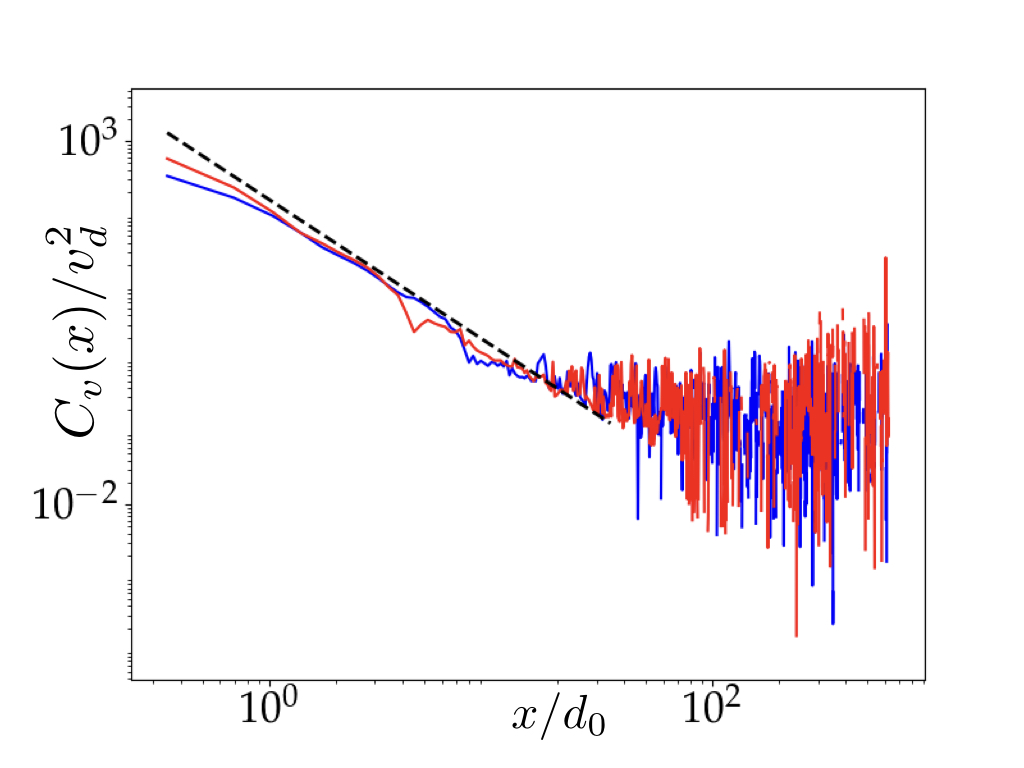}\includegraphics[width=.5\linewidth]{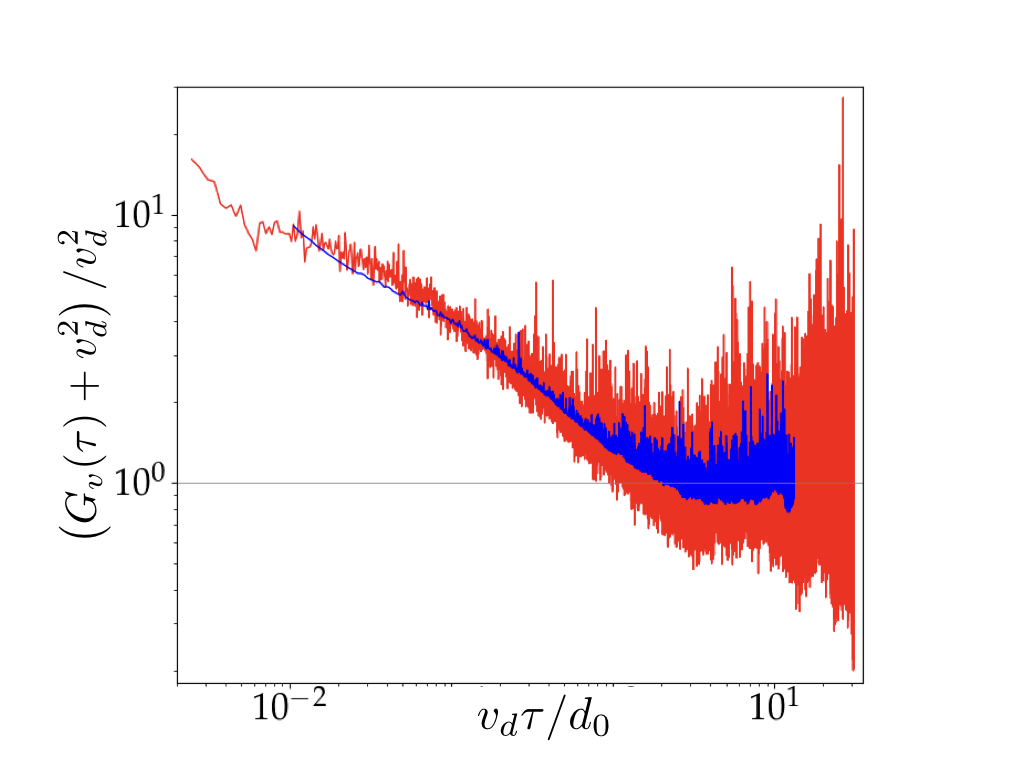}
\caption{Raw signal for the spatial (left) and temporal (right) correlations in the experiment for driving velocities 
$v_1 = 132$ nm/s (blue) and $v_2=31$ nm/s (red).
\label{Fig:RawCorrelationFcts}}
\end{figure}
\begin{figure}[h]
\includegraphics[width=.5\linewidth]{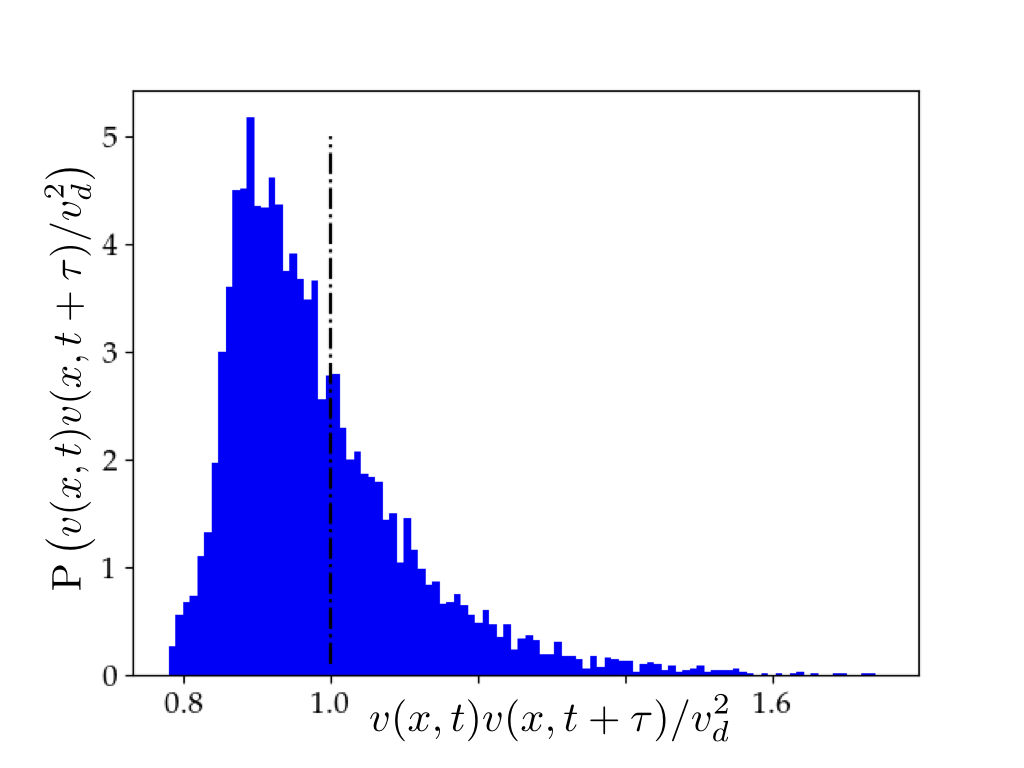}\includegraphics[width=.5\linewidth]{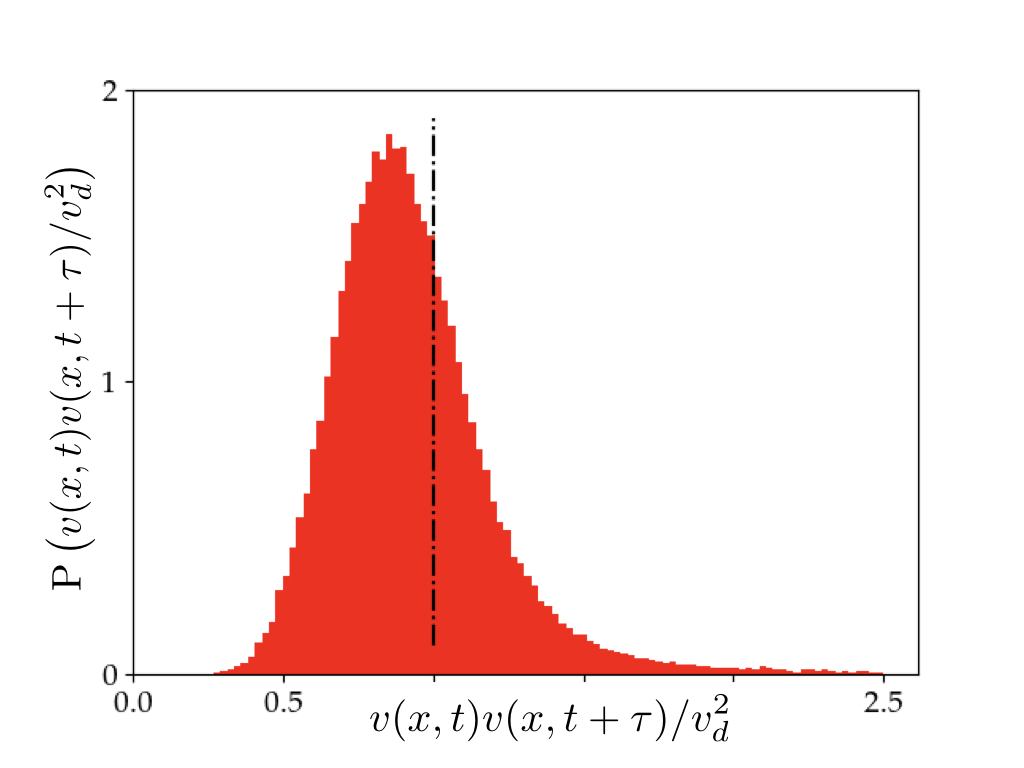}
\caption{\textbf{Left :} Histogram of $v(x,t)v(x,t+\tau)$ for driving velocity $v_1 = 132$ nm/s and for $\tau > 3$ in rescaled units (this corresponds to the point where we start to have some signal below $1$ on the right panel)
\textbf{Right :} Same histogram for driving velocity $v_2 = 31$ nm/s and for $\tau > 2$ in rescaled units.
\label{Fig:HistVelocity}}
\end{figure}

When we extract the correlation functions $C_v(x)$ and $G_v(\tau)$ from the experiment we first obtain a very noisy curve (see Fig. \ref{Fig:RawCorrelationFcts}). In the plots presented in the main text the curves are smooth because we took the average values over bins of equal logarithmic size. 
When one looks at the raw temporal correlation function on the right panel of Fig. \ref{Fig:RawCorrelationFcts} it is not obvious to see wether we really have anticorrelation or not at large times.
To discriminate we plot on Fig. \ref{Fig:HistVelocity} the histogram of  $v(x,t)v(x,t+\tau)/v_d^2$ for all $\tau$ large enough so that the function potentially has anticorrelation. The histograms are clearly peaked below $1$ for both driving velocities $v_1$ and $v_2$. The mean and median are respectively 0.983 and 0.951 for $v_1$ and 0.944 and 0.896  for $v_2$.
This shows that the anticorrelation is real and that the signal above 1 is due to the noise.

\end{widetext}

\end{document}